\documentclass{aa}
\usepackage{graphics,epsfig}
\usepackage{txfonts}
\usepackage{xcolor}

\def\simless{\mathbin{\lower 3pt\hbox
     {$\rlap{\raise 5pt\hbox{$\char'074$}}\mathchar"7218$}}}   
\def\simmore{\mathbin{\lower 3pt\hbox
     {$\rlap{\raise 5pt\hbox{$\char'076$}}\mathchar"7218$}}}   

\def\msun{~{\rm M}_\odot}

\begin{document}

   \title{On the origin of the hard X-ray tail in neutron-star X-ray
   binaries}

   \subtitle{}
  \author{
	P. Reig\inst{1,2}
	\and
  	N. Kylafis\inst{2,1}
          }

   \institute{IESL, Foundation for Research and Technology-Hellas, 71110, 
   		Heraklion, Greece \email{pau@physics.uoc.gr}
	 \and Physics Department \& Institute of Theoretical  \& 
                Computational Physics, University of Crete, 70013, 
   		Heraklion, Greece 
		\email{kylafis@physics.uoc.gr}
	}

 \authorrunning{Reig et al.}
\titlerunning{Hard tails in NSXB}

   \offprints{pau@physics.uoc.gr}
   
  \date{Received ; accepted}

  \abstract
{Neutron star X-ray binaries emit a compact, optically thick, relativistic
radio jet during low-luminosity, usually hard states, as Galactic
black-hole X-ray binaries do. When radio emission is bright, a hard
power-law tail without evidence for an exponential cutoff is observed in
most systems. }
{We have developed a jet model that explains many spectral and timing properties
of black-hole binaries in the states where a jet is present. Our goal is to
investigate whether our jet model can reproduce the hard tail, with
the correct range of photon index and the absence of a high-energy cutoff,
in neutron-star X-ray binaries.}
{We have performed Monte Carlo simulations of the Compton upscattering of
soft, accretion-disk or boundary layer photons, in the jet and computed the emergent 
energy spectra, as well as the time lag of hard photons with respect to 
softer ones as a function of Fourier frequency. 
We fit the energy spectra with a power law modified by an exponential cutoff at high
energy.}
{We demonstrate that our  jet model naturally explains the observed
power-law distribution with photon index in the range 1.8--3. With an
appropriate choice of the parameters, the cutoff expected from
Comptonization is shifted to energies above $\sim$300 keV, producing a pure
power law without any evidence for a rollover, in agreement with the
observations.}
{Our results reinforce the idea that the link between the outflow (jet) and
inflow (disk) in X-ray binaries does not depend on the nature of the
compact object, but on the process of accretion. Furthermore,
we address the differences of jets in black-hole and neutron-star 
X-ray binaries and  predict that 
the break frequency in the spectral energy distribution of
neutron-star X-ray binaries, as a class, will be lower than that of
black-hole binaries.}

\keywords{X-rays: binaries -- stars: neutron -- stars: binaries close 
               }

   \maketitle

\section{Introduction}

In black-hole (BHXB) and neutron-star (NSXB) X-ray binaries, the transfer
of mass from the optical companion to the compact star is mediated by an
optically thick, geometrically thin, accretion disk. Because the mass
to radius ratio in neutron stars and black holes is similar\footnote{A 5$\msun$
black hole has a Schwarzchild radius of $\sim$ 15 km.}, the characteristic
velocities and dynamical time scales near the compact object are also
similar ($v\sim 0.5c$, $t \simless 1$ ms). Thus, it is not
surprising that BHXB and NSXB show very similar  X-ray spectral and
temporal characteristics 
\citep{vanderklis94,maccarone03,vanderklis05,barret04,disalvo06,munoz14}.

It is generally accepted that  the X-ray spectral continuum of BHXB and
NSXB results from the sum of two main components: one thermal component
which follows a blackbody distribution and another non-thermal component
that is best described as a power law. The thermal component dominates at
lower energies, hence it is referred to as the soft component, while the
power-law component extends up to a few hundred keV and it is often known as
the "hard tail"  \citep{barret04,mclintock06,lin07,done07,church12}.

In BHXB, the soft thermal component is  attributed to the emission from
the innermost, hotter regions of the accretion disk, while the hard power
law results from Comptonization of low-energy photons by energetic
electrons.  The physical origin of the Comptonizing medium could be an
advection-dominated accretion flow \citep{narayan94,esin97}, a low angular
momentum accretion flow \citep{ghosh11,garain12}, or a radio jet
\citep{band86,georganopoulos02,reig03, giannios05, markoff05}.

In NSXB, the emission spectrum is more complex owing to the presence of a
hard surface. Thus below $\sim$ 10 keV, NSXB may display two soft
components: a single-temperature blackbody from the neutron star  surface
or from the boundary layer between the disk and the neutron star and a
multi-temperature blackbody from the accretion disk
\citep{mitsuda84,gilfanov03}. The temperature of the neutron star photons
is expected to be higher ($kT\sim 2-3$ keV) than that of the disk soft
photons ($kT\sim 1$ keV), because the effective area of the neutron star
surface is more compact than that of the accretion disk
\citep{farinelli07}. Above $\sim$ 20 keV, the most prominent spectral
feature in the X-ray spectrum of NSXB is a power-law distribution (the hard
tail), that usually extends  up to 200--300 keV without evidence for a
cutoff
\citep{disalvo00,disalvo01,iaria01,damico01,farinelli05,migliari07,ding11}.

Both kinds of binary systems exhibit distinct spectral states in response
to changes in the mass accretion rate, for example, in the course of an
outburst. The main of these states are generally referred to as soft  and
hard, depending on whether the bulk of the luminosity is radiated below or
above a few keV, respectively. 

Another common property to both BHXB and NSXB is the presence of radio jets
\citep{fender04,migliari06}. Optically thick, compact, steady radio
emission is detected during the low (in the 1-20 keV band) X-ray lumonisity
hard state, while optically thin radio flares occur during transitions from
the hard to the soft sate. The jet is quenched in the soft state, when
the X-ray luminosity approaches the Eddington limit
\citep{paizis06,migliari07,miller-jones10}.

A useful way to describe the rich phenomenology exhibited by these systems
in the X-ray band is the hardness-luminosity diagram (HLD). In the HLD, BHXB
and NSXB display distinct curves, along which the sources move smoothly
as the mass accretion rate varies. 
In Kylafis \& Belloni (2015), a physical inerpretation was offered for the
phenomenology of BHXB along the q-shaped curve in the HLD.
In NSXB, the shape of the curve in the HLD devides them 
into $Z$ sources and {\it atoll} sources
\citep{hasinger89,vanderklis06}. The three branches that form the Z-shaped
HLD are called horizontal (HB), normal (NB), and flaring (FB) branches.
The  HB corresponds to the hardest X-ray state, while the FB is the soft
state. In the case of atoll NSXB, the branches are called island and banana
branches, with subcategories such as extreme island and lower and upper
banana. The extreme island branch corresponds to the hardest spectral sate.
The hard tail has been detected in almost all the currently known Z
sources  \citep{disalvo02} and many of the atoll sources
\citep{paizis06}.

The association of the compact radio jet and the X-ray hard tail with a
specific region of the source in the HLD is well documented in BHXB
\citep{fender09} and NSXB \citep{paizis06,migliari06}. Both the radio
emission and the strength of the hard tail become weaker at higher
accretion rates. Radio and hard X-rays show the strongest intensity in the
hard states of BHXB, the horizontal branch of $Z$-NSXB
\citep{hjellming90b,migliari07,dai07}, and the island or lower banana
states in {\it atoll}-NSXB \citep{migliari03}. In BHXB, radio emission is
absent in the soft sate. In some NSXB, radio emission may be still detected
in the softer states, albeit highly reduced \citep{homan04}.

While there is a general consensus that the radio emission is produced by a
compact jet, the physical origin of the hard tail is not well understood. 
It naturally arises from Comptonization, however, the details of how the
electrons acquire their energy (thermal, non-termal, or  bulk motion)  and
the source of seed photons (neutron-star surface, inner accretion disk,
synchrotron photons) remain unclear
\citep{barret00,disalvo06,paizis06,lin07,markoff05,farinelli09,revnivtsev14}.
A weak hard tail contributing a small percentage of the total flux
($\simless 3$\%) is also detected in the soft state, when the jet is
quenched or absent \citep[see, e.g.,][and references therein]{mclintock06}.
In this state, the power-law tail is thought to come from Comptonization in
non-thermal flares above and below the thermal disk
\citep{Poutanen97,gierlinski99}.

In a series of papers
\citep{reig03,giannios04,giannios05,kylafis08,reig15}, we showed that
Compton upscattering of soft photons from the accretion disk in the jet can
explain a number of observational relations between the spectral and timing
parameters in the hard state of BHXB. Our results clearly demonstrate that 
jets play a central role in all the observed phenomena, not only in the
radio emission.  Motivated by the similarities of the X-ray spectral
continuum in BHXB and NSXB and {\it especially by the clear connection  of
the presence of a hard tail with a radio-loud state of the source}, we
investigate whether our model can also account for the spectral properties
of NSXB. 

Our objective in this paper is to demonstrate that 
emission from the neutron-star surface plus Comptonization in a jet
can reproduce the observed spectra of bright NSXB, namely a soft thermal
component that is well described by a blackbody distribution with $kT=2-3$
keV and a hard tail that follows a power-law with photon index in the range
1.8--3 with no evidence for a cutoff up to 200--300 keV.

\begin{figure}
\begin{center}
\includegraphics[width=8cm]{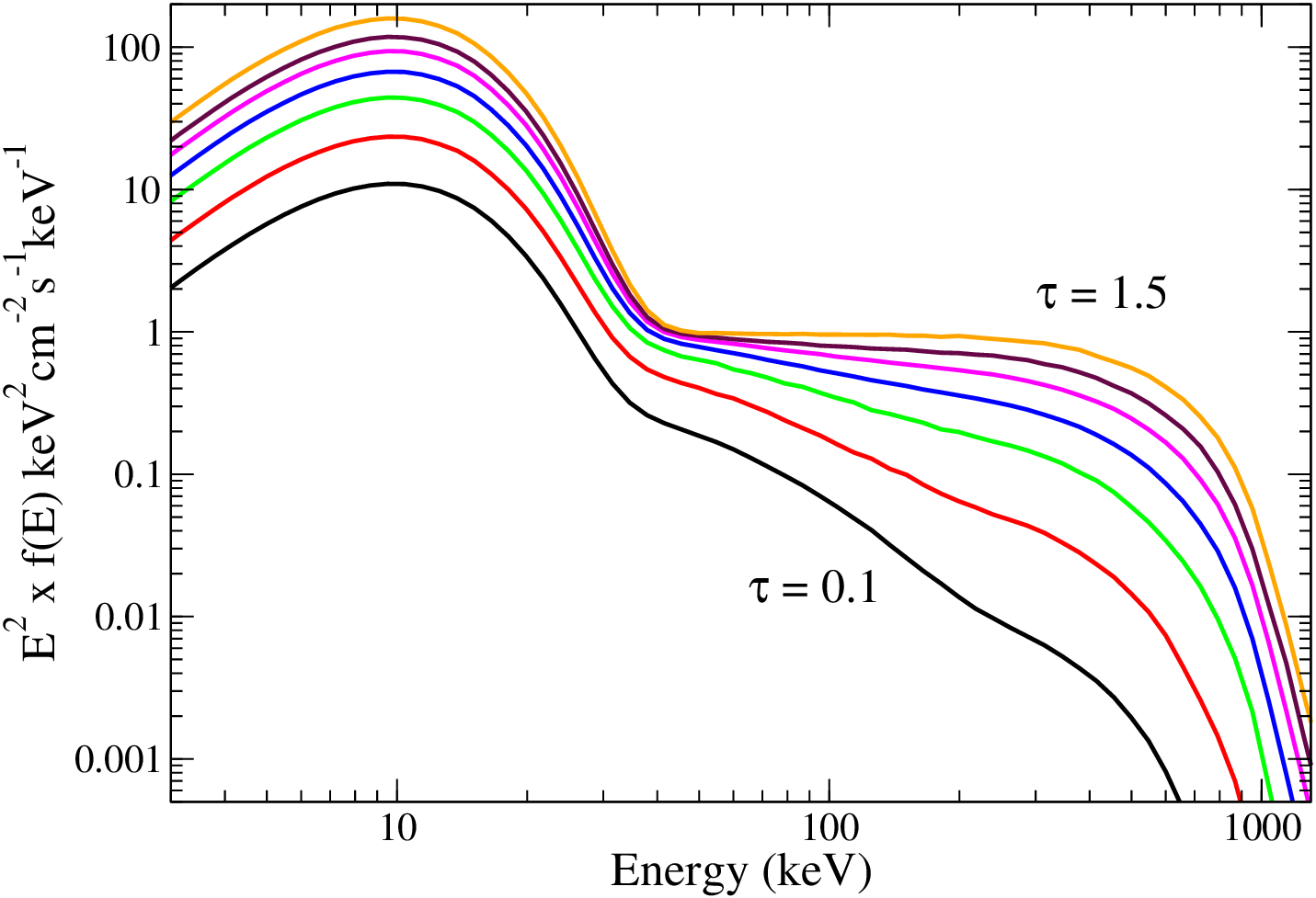}
\end{center}
\caption[]{Emergent spectra from the Monte Carlo simulations. The models
plotted correspond to a fixed $\gamma_{\rm min}=3.34$ ($v_{\perp}=0.52c$)
and optical depth $\tau=0.1,0.25,0.5,0.75,1.0,1.25,1.5$. The input source of photons 
follows a blackbody distribution with $kT=2.5$ keV. 
The normalization of the input blackbody
was chosen so that the hard component contributes 
about 5\% of the 0.1-300 keV luminosity. In the energy range covered by
current missions ($<300$ keV) the hard tail shows no break.}
\label{fig1}
\end{figure}

\section{The model}
\label{model}

The model that we have used  in this work is identical to that used in
\citet{reig15}.   For our Monte Carlo code we follow
\citet{pozdnyakov83}.  Photons from a blackbody distribution of
temperature $kT_{bb}$ are injected at the 
base of the jet with an upward isotropic distribution. As the photons
travel through the medium, they experience Compton scatterings with the
spiraling electrons. In each scattering, the photons gain 
on average energy from the bulk
motion (i.e. $v_{\parallel}$ and $v_{\perp}$) of the electrons.  
Comptonization can
occur everywhere in the jet. The optical depth to electron scattering, the
energy change and the new direction of the photons after scattering are
computed using the corresponding relativistic expressions. Each model is
run for $10^7$ photons. More details of how the code works can be found in
\citet{kylafis08}.

Since we are not interested in reproducing the radio
spectrum \citep[see][]{giannios05}, we have assumed mono-energetic
electrons in the jet with  Lorentz factor equal to the smallest in the
distribution, namely

\begin{equation}
\label{lorenzt}
\gamma_{\rm min}=\frac{1}{\sqrt{1-(v_{\parallel}^2+v_{\perp}^2)/c^2}},
\end{equation}

\noindent where $v_\parallel = v_0 =$ constant 
is the terminal velocity of the jet and $v_\perp$ is the 
smallest peprendicular velocity of the electrons in the lab frame.

The flow velocity in the jet is given by

\begin{equation}
\label{accel}
v_{\parallel}(z) =
\begin{cases}
v_0 ~ (z/z_1)^p  & \text{if } 0< z \leqslant z_1\\
v_0                   & \text{if } z> z_1,
\end{cases}
\end{equation}

\noindent where $z_1$ and $p$ are parameters.  In other words, the jet has an
acceleration region of thickness $z_1$, beyond which the flow has 
constant velocity $v_0$.

For a parabolic jet, i.e. one whose radius at height $z$ is 
$R(z)=R_0(z/z_0)^{1/2}$, the electron density is obtained from the
continuity equation and for $z>z_1$ it is inversely proportional to $z$.

The fixed parameters of our models and their reference values are:   the
radius $R_0 = 50 \, R_{\rm NS}$ of the base of the jet, where  $R_{\rm
NS}=1.25 \times 10^6$ cm is the radius of the neutron star,  the distance
$z_0 = 1 \, R_{\rm NS}$ of the bottom of the jet from the neutron
star center, the
height $H=10^5 \, R_{\rm NS}$ of the jet, the terminal velocity $v_0= 0.8
c$ of the jet,  the thickness $z_1= 5 \, R_{\rm NS}$ of the acceleration
zone, the exponent $p=1/2$, and the temperature  $kT_{bb} = 2.5$ keV of the
soft-photon input.

The parameters of our model that we have varied are the Thomson optical
depth $\tau_{\parallel}$ along the axis of the jet and the  minimum Lorentz
factor $\gamma_{\rm min}$. Since $v_0$ is  a constant in our models, the
variation of $\gamma_{\rm min}$ is equivalent to a variation in
$v_{\perp}$.

Because the jet is relativistic, the results also depend on the angle
$\theta$ of observation with respect to the jet axis.  Here we consider an
intermediate range of observing angles $0.2 < \cos \theta < 0.6$.
Practically, for the Monte Carlo simulation this means that we count only
photons that leave the jet in this range of angles.

 \begin{figure}[!t]
   \centering
   \includegraphics[width=8cm]{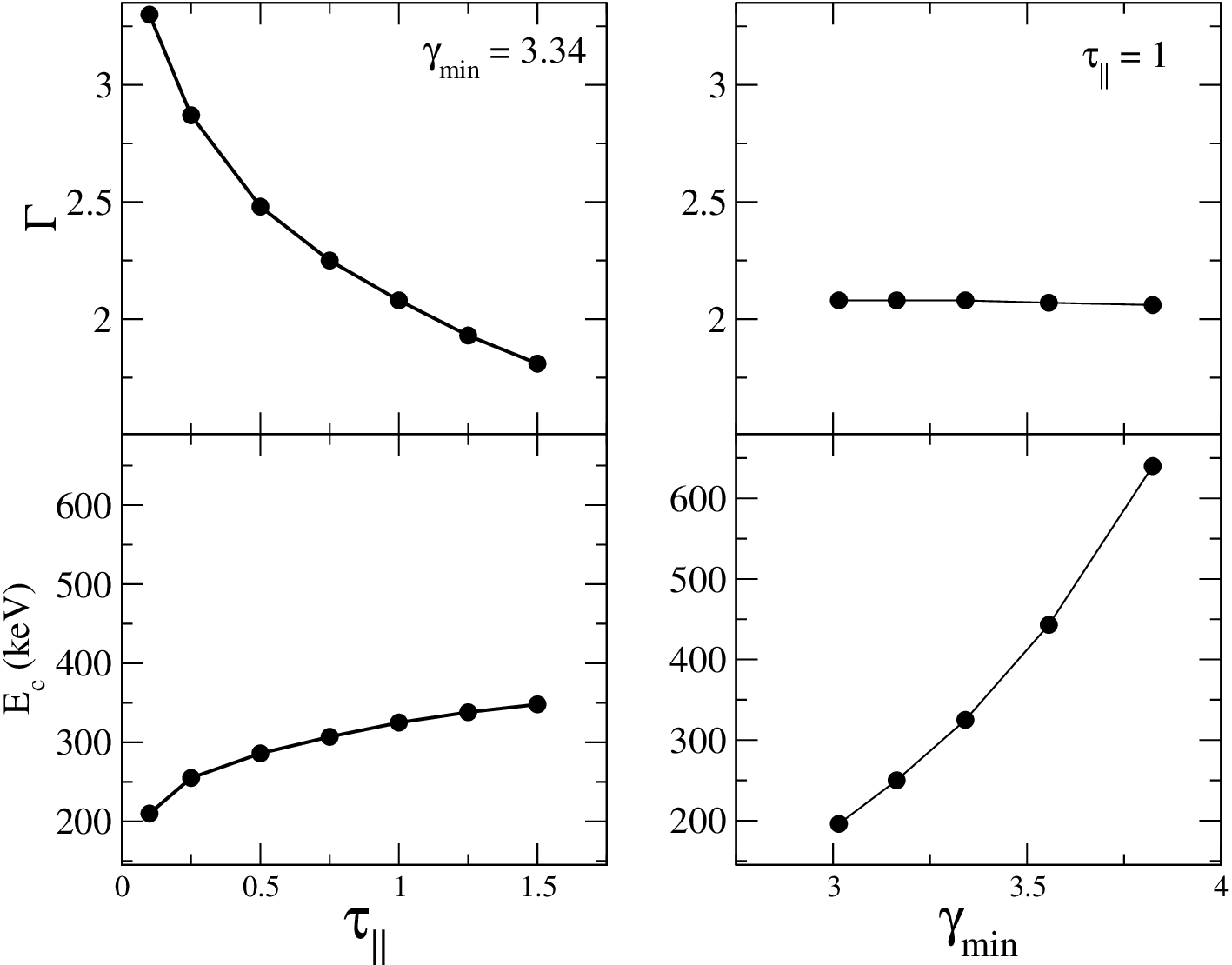}
   \caption{Photon index and cutoff energy as functions of $\tau_{\parallel}$ 
   and $\gamma_{\rm min}$. On the left panels, each point represents a 
   calculation with the same $\gamma_{\rm min}=3.34$ ($v_{\perp}=0.52c$) 
   and different $\tau_{\parallel}$.  On the right panels, each point 
   represents a calculation with the same $\tau_{\parallel}=1$ and different
   $\gamma_{\rm min}$. The scale of the Y-axis was left the same to
   faciliate the comparison.}
   \label{fig2}
 \end{figure}

\begin{table}
\caption{Results of our Monte Carlo simulations. Different models
correspond to different values of the optical depth ($\tau_{\parallel}$)
and perpendicular component of the electron velocity ($\gamma_{\rm min}$).
All the remaining parameters of the model were fixed at the reference
values (see Sect.~\ref{model}).}
\label{simu}
\begin{center}
\begin{tabular}{cccc}
\hline	\hline
$\tau_{\parallel}$     &$\gamma_{\rm min}$/$v_{\perp}$ (c)     &$\Gamma$    &$E_{\rm c}$ (keV)       \\        
\hline
0.10	&3.015/0.50    &3.30	&133	 \\
0.10	&3.164/0.51    &3.31	&172	 \\
0.10	&3.341/0.52    &3.28	&223	 \\
0.10	&3.556/0.53    &3.26	&314	 \\
0.10	&3.824/0.54    &3.27	&571	 \\
0.25	&3.015/0.50    &2.82	&151	 \\
0.25	&3.164/0.51    &2.82	&195	 \\
0.25	&3.341/0.52    &2.81	&253	 \\
0.25	&3.556/0.53    &2.79	&355	 \\
0.25	&3.824/0.54    &2.79	&566	 \\
0.50	&3.015/0.50    &2.44	&173	 \\
0.50	&3.164/0.51    &2.44	&219	 \\
0.50	&3.341/0.52    &2.43	&288	 \\
0.50	&3.556/0.53    &2.42	&392	 \\
0.50	&3.824/0.54    &2.41	&588	 \\
0.75	&3.015/0.50    &2.21	&187	 \\
0.75	&3.164/0.51    &2.20	&238	 \\
0.75	&3.341/0.52    &2.20	&309	 \\
0.75	&3.556/0.53    &2.19	&422	 \\
0.75	&3.824/0.54    &2.18	&604	 \\
1.00	&3.015/0.50    &2.01	&195	 \\
1.00	&3.164/0.51    &2.02	&251	 \\
1.00	&3.341/0.52    &2.01	&322	 \\
1.00	&3.556/0.53    &2.01	&438	 \\
1.00	&3.824/0.54    &2.00	&628	 \\
1.25	&3.015/0.50    &1.88	&205	 \\
1.25	&3.164/0.51    &1.89	&262	 \\
1.25	&3.341/0.52    &1.89	&344	 \\
1.25	&3.556/0.53    &1.88	&461	 \\
1.25	&3.824/0.54    &1.88	&665	 \\
1.50	&3.015/0.50    &1.73	&210	 \\
1.50	&3.164/0.51    &1.74	&268	 \\
1.50	&3.341/0.52    &1.74	&347	 \\
1.50	&3.556/0.53    &1.73	&457	 \\
1.50	&3.824/0.54    &1.73	&640	 \\
\hline
\hline
\end{tabular}
\end{center}
\end{table}

\section{Results}
\label{results}

The spectral parameters relevant to the present study are the photon index
of the hard-tail component and the cutoff energy. We have investigated the
dependence of these two quantities on the choice of two parameters of our
model, namely the optical depth and the electron velocity (see
Table~\ref{simu}).

Figure \ref{fig1} shows the emerging spectra that result from our
simulations using the reference values of the parameters reported in
Sect.~\ref{model} with $\gamma_{\rm min}=3.34$ ($v_{\perp}=0.52c$) and
various optical depths $\tau_{\parallel}$ =  0.1, 0.25, 0.5, 0.75, 1.0,
1.25, and 1.5. As it can be seen, although the spectra roll over above a
certain energy, this energy is well above 200-300 keV. Thus, up to 200 -
300 keV, the spectra are essentially pure power-laws with photon index
between 1.8--3, as observed. Figure \ref{fig2} shows the
photon index $\Gamma$ and the cutoff energy $E_{\rm c}$ as functions of
optical depth $\tau_{\parallel}$ and $\gamma_{\rm min}$ (or equivalently
$v_{\perp}$). The scale of the  vertical axis was kept the same in the two
plots to better assess the differences. The cutoff energy has a weak
dependence on $\tau_{\parallel}$ for the range of optical depths
considered, but it is strongly dependent of $\gamma_{\rm min}$ . This is
expected because the cutoff is mainly determined by the energetics of the
electrons \citep{giannios05}. In contrast, the photon index strongly
depends on $\tau_{\parallel}$, but only weakly on  $\gamma_{\rm min}$. The
reason is that the slope of the spectrum is very sensitive to the number of
scatterings, which increases with optical depth.

Because of the significant flow velocity in the jet, the up-scattered
photons escape preferentially in the forward direction. Thus we expect the
results to depend also on the angle $\theta$ of observation with respect to
the jet axis. The hardest spectra are found at small observation angles, as
these are the photons that have suffered the most energetic scatterings. The
cutoff energy increases as the observation angle decreases, but stabilizes
at intermediate angles. However, once the angle range is fixed, the
parameters follow a similar trend with optical depth and Lorentz gamma
factor, irrespective of the range considered.

\begin{figure}[!t]
   \centering
   \includegraphics[width=8cm]{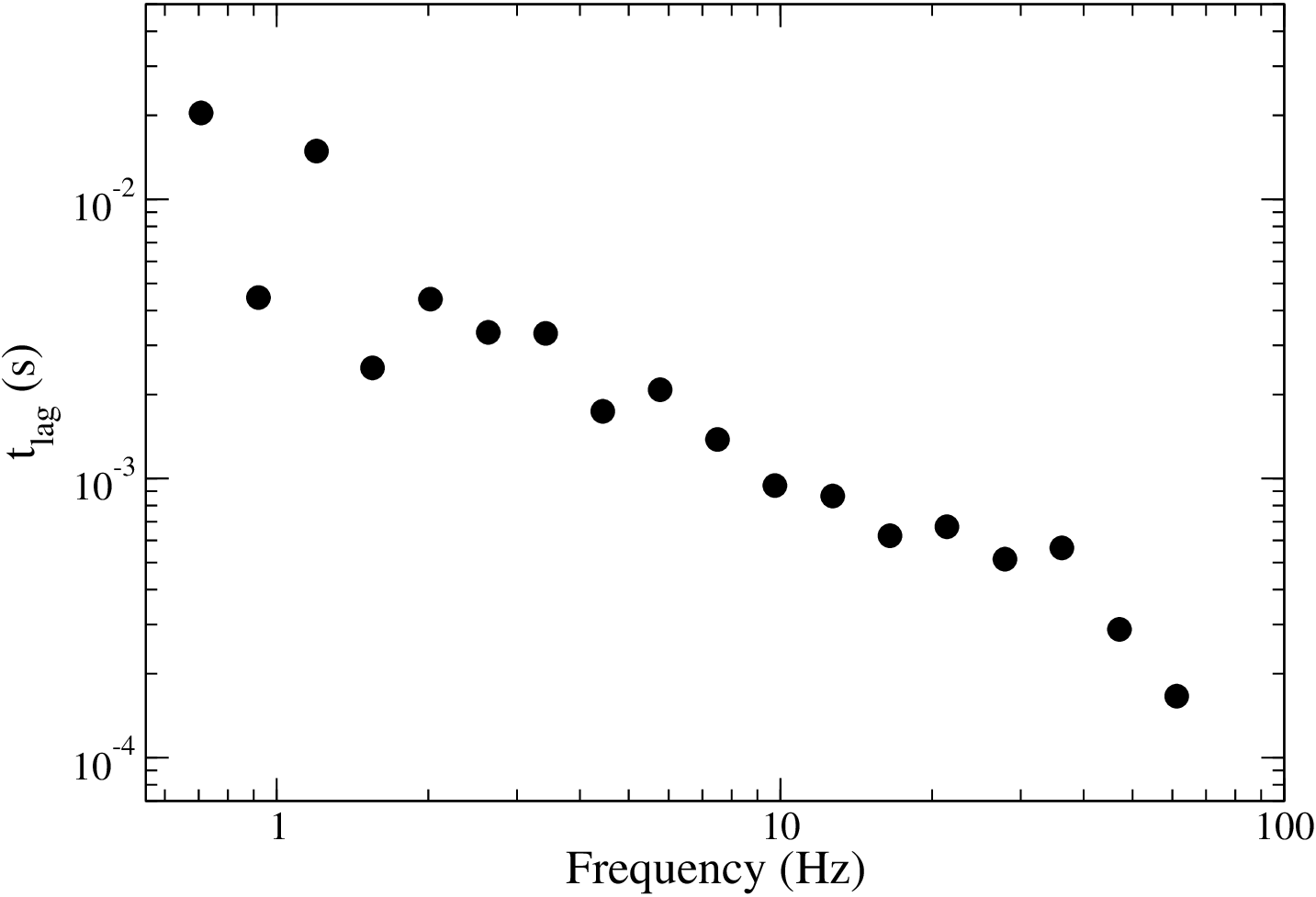}
   \caption{Time-lag spectrum that results from the model with 
  $\tau_{\parallel}=1$ and $\gamma_{\rm min}=3.34$. All the remaining parameters 
  of the model were fixed at the reference values (see Sect.~\ref{model}).}
   \label{fig3}
 \end{figure}

We have also computed the time lag of hard photons in the energy range
7--40 keV with respect to softer photons in the range 2--7 keV. The
time-lag spectrum, i.e. the time lag as a function of Fourier frequency, is
shown in Fig.~\ref{fig3}. The spectra follow a power-law dependence on
Fourier  frequency $t_{\rm lag} \sim \nu^{-\beta}$,  with $\beta =
1.0\pm0.1$. This slope is somewhat steeper than the average slope measured
in black-hole binaries, $t_{\rm lag}\sim \nu^{-\alpha}$ with $\alpha =
0.7\pm0.1$ \citep{nowak99,pottschmidt03,cassatella12}, but agrees with the
results reported by \citet{olive01}.  We have not found a significant
dependence of the shape of the lag spectrum with either the optical depth
or $\gamma_{\rm min}$. 

\section{Discussion}

We have run Monte Carlo simulations of the Compton upscattering of soft 
photons in an outflow moving at relativistic velocity (representing a radio
jet) and computed the emerging X-ray spectrum. The input parameters of the
model have been selected so that the results  match what is seen in
bright NSXB.  Assuming a blackbody input source of photons with $kT$=2.5
keV, our model reproduces very well the observed 1-300 keV X-ray spectrum
in the harder states, a power law continuum with a photon index $\Gamma$ in
the range 1.8--3 without evidence for a cutoff up to 200--300 keV.

The main difficulty of the models that seek to reproduce the 
high-energy X-ray spectral
continuum in NSXB lies in explaining the absence of a cutoff, because the
spectrum resulting from Comptonization is expected to roll over at a certain
energy (an effect known as Compton recoil), when the energy of the photons
approaches that of the electrons. Various models have been put forward to
explain the absence of a cutoff in the hard tails: bulk motion
Comptonization \citep{bradshaw03,paizis06,farinelli07,farinelli08}, Comptonization by a
hybrid--thermal-non thermal corona
\citep{gierlinski99,ozel00,zdziarski01,farinelli05,revnivtsev14}, or
synchrotron emission from the electrons of a jet \citep{markoff05}. Because
of the limited sensitivity of current detectors at energies above $\sim$100
keV, statistically significant detections are given up to $\sim$200 keV,
typically. The spectral continuum could in principle still roll over at a
certain energy provided that this cutoff energy is larger than the limit
energy of the detectors\footnote{From the point of view of the energetics,
if the slope of the power law is $\Gamma \leq 2$, then the spectrum must
show a rollover at certain energy to prevent a diverging luminosity.}. Recently,
\citet{revnivtsev14} reported the highest lower limit on the cutoff energy
that has been detected so far: $E_{\rm c}> 330$ keV, at a  $2\sigma$
significance level.

This relatively high lower limit of the cutoff energy was used by
\citet{revnivtsev14} to rule out the bulk-motion Comptonization model. The
reason is that the cutoff energy in thermal and bulk-motion Comptonization
depends on the temperature and the velocity of the Comtonizing electrons,
respectively. For realistic values of $kT_e < 50$ keV and assuming
free-fall velocity $v_{\rm ff} \approx 0.5 c$
onto a neutron star of canonical mass ($1.4\msun$) and
radius ($10^6$ cm), the cutoff energy is expected
to be below 200 keV. Moreover, the radiation pressure expected to result
from the emission of
the neutron-star surface may reduce the infall velocity to less than
$0.2c$, decreasing the bulk motion Comptonization effect \citep{farinelli08}.

Observations collected over the past decade show that there exists a clear
connection between the X-ray spectral states and the radio emission in
NSXB. Similarly to BHXB,  the NSXB are radio loud in the hard state,
usually displaying a flat spectrum indicative of a relativistic jet. The
radio emission is quenched when the source enters the soft state in most
(but not all) cases
\citep{penninx88,hasinger90,hjellming90b,oosterbroek94,migliari06,migliari07,miller-jones10}.

Motivated by the disputed origin of the hard X-ray tail in NSXB and by the
clear relationship between the presence of this hard component and  radio
emission from the source in the form of a jet, we have investigated whether
the observed X-ray spectrum of NSXB can be reproduced by our jet model. Our
goal has been to reproduce the observed spectrum above $\sim$20 keV,  
namely, a power-law distribution with a photon index in the range 1.8--3
and the absence of a high-energy cutoff up to an observed energy of  300
keV. In Sect.~\ref{results}, we demonstrated that such a range of photon
indices can be obtained with an optical depth varying in the range $0.1
\le \tau_{\parallel} \le 1.5$ (Table~\ref{simu}). The lack of a cutoff in
the observed spectra translates into cutoff energies above 200 keV in our
models. These energies are obtained when the electrons move with
$\gamma_{\rm min} \simmore 3.2$  ($v_{\perp} \simmore 0.52 c$, for
$v_{\parallel} = 0.8 c$). Note that, in principle, and contrary to the bulk
motion Comptonization model, there is no restriction on the velocity
components of the electrons in the jet, provided that the parallel
component is significantly larger than the perpendicular component.
Therefore, although we selected models that give cutoff energies above but
not too far from 300 keV, we could increase our limit on $E_{\rm c}$ simply
by increasing $\gamma_{\rm min}$ (see Fig~\ref{fig2}).  Note that such high
$\gamma_{\rm min}$ is possible, since \citet{fomalont01b}, who studied the
evolution of the radio emission in  Sco X--1, found that  $v > 0.95 c$.

\subsection{Comparison with black-hole binaries}

There are two major differences between the jets in NSXB and BHXB: {\em i)}
the radio luminosities of neutron-star jets are typically $\sim$30 times
lower than those of black-hole jets at comparable X-ray luminosities
\citep{fender01, migliari03} and {\em ii)} radio emission is not completely
quenched in the soft state of some NSXB \citep{migliari04}, as opposed to
their non-detection in BHXB, when they are in the same state.

Both of these differences have been addressed successfully in
\citet{kylafis12}, where the formation and the destruction of jets is
explained using the Cosmic Battery
\citep{contopoulos98,contopoulos06,christodoulou08}.

That the radio luminosity is significantly lower in neutron-star
jets than in black-hole ones can be understood as follows:
while the radio luminosity scales as $\dot m M$ for both types
of compact object (here, $\dot m$ is the mass accretion rate
in units of the Eddington value and $M$ is the mass of the compact object),
the X-ray luminosity 
due to the ADAF-like inner accretion flow scales as $\dot m M$ for 
neutron stars and as $\dot m^2 M$ for black holes 
\citep{narayan97, migliari06, abramowicz13}.
Thus, for comparable X-ray luminosities, the radio luminosity of 
neutron-star jets is $\sim \dot m$ times smaller than that of black-hole ones. 

The radio emission is quenched in the soft state of BHXB, because the
Cosmic Battery works very inefficiently when the accretion flow is in the
form of a Shakura-Sunyayev type \citep{shakura73}.  On the other hand, in
low-mass NSXB with weak magnetic field ($B\sim 10^{8}-10^{9}$ G), the
Cosmic Battery can work moderately  efficiently due to the ``spreading
layer'' \citep{inogamov10} that forms on the surface of the neutron star,
provided that there is  a significant difference between the spin frequency
of the neutron star  and the Keplerian frequency of the disk. 

The findings from our calculations are consistent with the characteristics
of neuron-star and black-hole jets discussed above.  We have found that
in NSXB the required optical depths are in the range  $0.1 \le
\tau_{\parallel} \le 1.5$, while in BHXB they are in the range $1 \le
\tau_{\parallel} \le 10$ \citep{reig03,kylafis08, reig15}. Because the
optical depth is proportional to density and so is the radio emission, the
lower optical depth in NSXB is consistent with the observational fact that
the radio emission of the jet in NSXB is significantly smaller than that in
BHXB.

To explain the fact that BHXB exhibit a high-energy cutoff in the X-ray
spectrum while NSXB do not, we were forced to assume that $\gamma_{\rm
min}$ in the jets of NSXB is larger than in BHXB.   In particular, we found
from our calculations that for BHXB  $\gamma_{\rm min} \simless 2.4$
\citep{reig15}, while for NSXB $\gamma_{\rm min} \simmore 3$.  It will be
interesting to see if this  is confirmed by observations.

In a recent paper \citep{koljonen15}, an important  anti-correlation was
found between the break frequency $\nu_b$ in the jet spectrum (spectral
energy distribution) and the power-law index $\Gamma$ of the hard X-ray
spectrum for BHXB and AGN.  If the X-ray spectrum is produced in the jet by
inverse Compton scattering \citep{reig03, giannios04, giannios05,
kylafis08, reig15}, then this anticorrelation can be understood at least
qualitatively.  As the jet weakens, $\tau_{\parallel}$ decreases, and as a
result  $\Gamma$ increases \citep{reig15} and $\nu_b$ decreases. If our
model is correct, i.e., the hard X-ray spectrum is produced in the jet by
inverse Comptonization, then we {\em predict that NSXB as a class will
exhibit lower values of $\nu_b$ than BHXB}.

In our model, time or phase lags result from the random walk of the
photons in the scattering medium. Hard photons scatter on average
more times than softer ones
before escaping the Comptonizing medium, i.e. the jet, hence they
are delayed with respect to the softer ones. While the amplitude of the lags
depends on the optical depth and the dimensions of the jet
\citep{giannios04,kylafis08}, the slope of the time-lag spectrum
appears to be rather insensitive to those parameters or the Lorentz factor.
In practice, the slope is determined for given energy ranges of the soft
and hard photons and calculated for a given frequency range. Although most of
our models produce slopes $\sim$ 1, smaller indices, similar to those of
BHXB, could be obtained by different combinations of frequency and energy
ranges. Given the lack of publications reporting Fourier time-lag spectra
of NSXB, we cannot conclusively establish whether the different slopes that
our calculations suggest corresponds to a distinguishing feature between
NSXB and BHXB.

\section{Conclusion}

We have run Monte Carlo simulations of the Compton upscattering of soft 
photons in an outflow moving at relativistic velocity (representing a radio
jet) and computed the emerging X-ray spectrum of bright NSXB.   Our Monte
Carlo simulations reproduce the observed spectral features:  a power-law
energy distribution at energies above 20 keV with photon index in the range
1.8--3 and the absence of a high energy cutoff  up to $\sim$300 keV. We
demonstrate that the hard tail detected in these syetms can be explained by
Comptonization in the jet. The observed connection between the hard X-ray
tail and the radio emission in low-mass X-ray binaries with neutron star
companions supports this scenario.  We explain the differences in the
spectral and timing parameters between BHXB and NSXB by assuming that the
electron population in the jets of NSXB have higher velocities and lower
densities than in BHXB.

\begin{acknowledgements}

This research has been supported in part by the
"RoboPol" project, which is implemented under the "ARISTEIA" Action of the
"OPERATIONAL PROGRAM EDUCATION AND LIFELONG LEARNING" and is co-funded by
the European Social Fund (ESF) and National Resources.

\end{acknowledgements}

\bibliographystyle{aa}
\bibliography{../art-NS}

\begin{thebibliography}{73}
\expandafter\ifx\csname natexlab\endcsname\relax\def\natexlab#1{#1}\fi

\bibitem[{{Abramowicz} \& {Fragile}(2013)}]{abramowicz13}
{Abramowicz}, M.~A. \& {Fragile}, P.~C. 2013, Living Reviews in Relativity, 16,
  1

\bibitem[{{Band} \& {Grindlay}(1986)}]{band86}
{Band}, D.~L. \& {Grindlay}, J.~E. 1986, \apj, 311, 595

\bibitem[{{Barret}(2004)}]{barret04}
{Barret}, D. 2004, in American Institute of Physics Conference Series, Vol.
  703, Plasmas in the Laboratory and in the Universe: New Insights and New
  Challenges, ed. G.~{Bertin}, D.~{Farina}, \& R.~{Pozzoli}, 238--249

\bibitem[{{Barret} {et~al.}(2000){Barret}, {Olive}, {Boirin}, {Done},
  {Skinner}, \& {Grindlay}}]{barret00}
{Barret}, D., {Olive}, J.~F., {Boirin}, L., {et~al.} 2000, \apj, 533, 329

\bibitem[{{Bradshaw} {et~al.}(2003){Bradshaw}, {Geldzahler}, \&
  {Fomalont}}]{bradshaw03}
{Bradshaw}, C.~F., {Geldzahler}, B.~J., \& {Fomalont}, E.~B. 2003, \apj, 592,
  486

\bibitem[{{Cassatella} {et~al.}(2012){Cassatella}, {Uttley}, {Wilms}, \&
  {Poutanen}}]{cassatella12}
{Cassatella}, P., {Uttley}, P., {Wilms}, J., \& {Poutanen}, J. 2012, \mnras,
  422, 2407

\bibitem[{{Christodoulou} {et~al.}(2008){Christodoulou}, {Contopoulos}, \&
  {Kazanas}}]{christodoulou08}
{Christodoulou}, D.~M., {Contopoulos}, I., \& {Kazanas}, D. 2008, \apj, 674,
  388

\bibitem[{{Church} {et~al.}(2012){Church}, {Gibiec},
  {Ba{\l}uci{\'n}ska-Church}, \& {Jackson}}]{church12}
{Church}, M.~J., {Gibiec}, A., {Ba{\l}uci{\'n}ska-Church}, M., \& {Jackson},
  N.~K. 2012, \aap, 546, A35

\bibitem[{{Contopoulos} \& {Kazanas}(1998)}]{contopoulos98}
{Contopoulos}, I. \& {Kazanas}, D. 1998, \apj, 508, 859

\bibitem[{{Contopoulos} {et~al.}(2006){Contopoulos}, {Kazanas}, \&
  {Christodoulou}}]{contopoulos06}
{Contopoulos}, I., {Kazanas}, D., \& {Christodoulou}, D.~M. 2006, \apj, 652,
  1451

\bibitem[{{D'A{\'{\i}}} {et~al.}(2007){D'A{\'{\i}}}, {{\.Z}ycki}, {Di Salvo},
  {Iaria}, {Lavagetto}, \& {Robba}}]{dai07}
{D'A{\'{\i}}}, A., {{\.Z}ycki}, P., {Di Salvo}, T., {et~al.} 2007, \apj, 667,
  411

\bibitem[{{D'Amico} {et~al.}(2001){D'Amico}, {Heindl}, {Rothschild}, \&
  {Gruber}}]{damico01}
{D'Amico}, F., {Heindl}, W.~A., {Rothschild}, R.~E., \& {Gruber}, D.~E. 2001,
  \apjl, 547, L147

\bibitem[{{Di Salvo} {et~al.}(2002){Di Salvo}, {Farinelli}, {Burderi},
  {Frontera}, {Kuulkers}, {Masetti}, {Robba}, {Stella}, \& {van der
  Klis}}]{disalvo02}
{Di Salvo}, T., {Farinelli}, R., {Burderi}, L., {et~al.} 2002, \aap, 386, 535

\bibitem[{{Di Salvo} {et~al.}(2006){Di Salvo}, {Iaria}, {Robba}, \&
  {Burderi}}]{disalvo06}
{Di Salvo}, T., {Iaria}, R., {Robba}, N., \& {Burderi}, L. 2006, Chinese
  Journal of Astronomy and Astrophysics Supplement, 6, 183

\bibitem[{{Di Salvo} {et~al.}(2001){Di Salvo}, {Robba}, {Iaria}, {Stella},
  {Burderi}, \& {Israel}}]{disalvo01}
{Di Salvo}, T., {Robba}, N.~R., {Iaria}, R., {et~al.} 2001, \apj, 554, 49

\bibitem[{{Di Salvo} {et~al.}(2000){Di Salvo}, {Stella}, {Robba}, {van der
  Klis}, {Burderi}, {Israel}, {Homan}, {Campana}, {Frontera}, \&
  {Parmar}}]{disalvo00}
{Di Salvo}, T., {Stella}, L., {Robba}, N.~R., {et~al.} 2000, \apjl, 544, L119

\bibitem[{{Ding} {et~al.}(2011){Ding}, {Zhang}, {Wang}, {Qu}, \&
  {Yan}}]{ding11}
{Ding}, G.~Q., {Zhang}, S.~N., {Wang}, N., {Qu}, J.~L., \& {Yan}, S.~P. 2011,
  \aj, 142, 34

\bibitem[{{Done} {et~al.}(2007){Done}, {Gierli{\'n}ski}, \& {Kubota}}]{done07}
{Done}, C., {Gierli{\'n}ski}, M., \& {Kubota}, A. 2007, \aapr, 15, 1

\bibitem[{{Esin} {et~al.}(1997){Esin}, {McClintock}, \& {Narayan}}]{esin97}
{Esin}, A.~A., {McClintock}, J.~E., \& {Narayan}, R. 1997, \apj, 489, 865

\bibitem[{{Farinelli} {et~al.}(2005){Farinelli}, {Frontera}, {Zdziarski},
  {Stella}, {Zhang}, {van der Klis}, {Masetti}, \& {Amati}}]{farinelli05}
{Farinelli}, R., {Frontera}, F., {Zdziarski}, A.~A., {et~al.} 2005, \aap, 434,
  25

\bibitem[{{Farinelli} {et~al.}(2009){Farinelli}, {Paizis}, {Landi}, \&
  {Titarchuk}}]{farinelli09}
{Farinelli}, R., {Paizis}, A., {Landi}, R., \& {Titarchuk}, L. 2009, \aap, 498,
  509

\bibitem[{{Farinelli} {et~al.}(2007){Farinelli}, {Titarchuk}, \&
  {Frontera}}]{farinelli07}
{Farinelli}, R., {Titarchuk}, L., \& {Frontera}, F. 2007, \apj, 662, 1167

\bibitem[{{Farinelli} {et~al.}(2008){Farinelli}, {Titarchuk}, {Paizis}, \&
  {Frontera}}]{farinelli08}
{Farinelli}, R., {Titarchuk}, L., {Paizis}, A., \& {Frontera}, F. 2008, \apj,
  680, 602

\bibitem[{{Fender} {et~al.}(2004){Fender}, {Belloni}, \& {Gallo}}]{fender04}
{Fender}, R.~P., {Belloni}, T.~M., \& {Gallo}, E. 2004, \mnras, 355, 1105

\bibitem[{{Fender} {et~al.}(2009){Fender}, {Homan}, \& {Belloni}}]{fender09}
{Fender}, R.~P., {Homan}, J., \& {Belloni}, T.~M. 2009, \mnras, 396, 1370

\bibitem[{{Fender} \& {Kuulkers}(2001)}]{fender01}
{Fender}, R.~P. \& {Kuulkers}, E. 2001, \mnras, 324, 923

\bibitem[{{Fomalont} {et~al.}(2001){Fomalont}, {Geldzahler}, \&
  {Bradshaw}}]{fomalont01b}
{Fomalont}, E.~B., {Geldzahler}, B.~J., \& {Bradshaw}, C.~F. 2001, \apjl, 553,
  L27

\bibitem[{{Garain} {et~al.}(2012){Garain}, {Ghosh}, \&
  {Chakrabarti}}]{garain12}
{Garain}, S.~K., {Ghosh}, H., \& {Chakrabarti}, S.~K. 2012, \apj, 758, 114

\bibitem[{{Georganopoulos} {et~al.}(2002){Georganopoulos}, {Aharonian}, \&
  {Kirk}}]{georganopoulos02}
{Georganopoulos}, M., {Aharonian}, F.~A., \& {Kirk}, J.~G. 2002, \aap, 388, L25

\bibitem[{{Ghosh} {et~al.}(2011){Ghosh}, {Garain}, {Giri}, \&
  {Chakrabarti}}]{ghosh11}
{Ghosh}, H., {Garain}, S.~K., {Giri}, K., \& {Chakrabarti}, S.~K. 2011, \mnras,
  416, 959

\bibitem[{{Giannios}(2005)}]{giannios05}
{Giannios}, D. 2005, \aap, 437, 1007, "Paper III"

\bibitem[{{Giannios} {et~al.}(2004){Giannios}, {Kylafis}, \&
  {Psaltis}}]{giannios04}
{Giannios}, D., {Kylafis}, N.~D., \& {Psaltis}, D. 2004, \aap, 425, 163, "Paper
  II"

\bibitem[{{Gierli{\'n}ski} {et~al.}(1999){Gierli{\'n}ski}, {Zdziarski},
  {Poutanen}, {Coppi}, {Ebisawa}, \& {Johnson}}]{gierlinski99}
{Gierli{\'n}ski}, M., {Zdziarski}, A.~A., {Poutanen}, J., {et~al.} 1999,
  \mnras, 309, 496

\bibitem[{{Gilfanov} {et~al.}(2003){Gilfanov}, {Revnivtsev}, \&
  {Molkov}}]{gilfanov03}
{Gilfanov}, M., {Revnivtsev}, M., \& {Molkov}, S. 2003, \aap, 410, 217

\bibitem[{{Hasinger} \& {van der Klis}(1989)}]{hasinger89}
{Hasinger}, G. \& {van der Klis}, M. 1989, \aap, 225, 79

\bibitem[{{Hasinger} {et~al.}(1990){Hasinger}, {van der Klis}, {Ebisawa},
  {Dotani}, \& {Mitsuda}}]{hasinger90}
{Hasinger}, G., {van der Klis}, M., {Ebisawa}, K., {Dotani}, T., \& {Mitsuda},
  K. 1990, \aap, 235, 131

\bibitem[{{Hjellming} {et~al.}(1990){Hjellming}, {Stewart}, {White}, {Strom},
  {Lewin}, {Hertz}, {Wood}, {Norris}, {Mitsuda}, {Penninx}, \& {van
  Paradijs}}]{hjellming90b}
{Hjellming}, R.~M., {Stewart}, R.~T., {White}, G.~L., {et~al.} 1990, \apj, 365,
  681

\bibitem[{{Homan} {et~al.}(2004){Homan}, {Wijnands}, {Rupen}, {Fender},
  {Hjellming}, {di Salvo}, \& {van der Klis}}]{homan04}
{Homan}, J., {Wijnands}, R., {Rupen}, M.~P., {et~al.} 2004, \aap, 418, 255

\bibitem[{{Iaria} {et~al.}(2001){Iaria}, {Burderi}, {Di Salvo}, {La Barbera},
  \& {Robba}}]{iaria01}
{Iaria}, R., {Burderi}, L., {Di Salvo}, T., {La Barbera}, A., \& {Robba}, N.~R.
  2001, \apj, 547, 412

\bibitem[{{Inogamov} \& {Sunyaev}(2010)}]{inogamov10}
{Inogamov}, N.~A. \& {Sunyaev}, R.~A. 2010, Astronomy Letters, 36, 848

\bibitem[{{Koljonen} {et~al.}(2015){Koljonen}, {Russell},
  {Fern{\'a}ndez-Ontiveros}, {Markoff}, {Russell}, {Miller-Jones}, {van der
  Horst}, {Bernardini}, {Casella}, {Curran}, {Gandhi}, \& {Soria}}]{koljonen15}
{Koljonen}, K.~I.~I., {Russell}, D.~M., {Fern{\'a}ndez-Ontiveros}, J.~A.,
  {et~al.} 2015, \apj, 814, 139

\bibitem[{{Kylafis} {et~al.}(2012){Kylafis}, {Contopoulos}, {Kazanas}, \&
  {Christodoulou}}]{kylafis12}
{Kylafis}, N.~D., {Contopoulos}, I., {Kazanas}, D., \& {Christodoulou}, D.~M.
  2012, \aap, 538, A5

\bibitem[{{Kylafis} {et~al.}(2008){Kylafis}, {Papadakis}, {Reig}, {Giannios},
  \& {Pooley}}]{kylafis08}
{Kylafis}, N.~D., {Papadakis}, I.~E., {Reig}, P., {Giannios}, D., \& {Pooley},
  G.~G. 2008, \aap, 489, 481, "Paper IV"

\bibitem[{{Lin} {et~al.}(2007){Lin}, {Remillard}, \& {Homan}}]{lin07}
{Lin}, D., {Remillard}, R.~A., \& {Homan}, J. 2007, \apj, 667, 1073

\bibitem[{{Maccarone} \& {Coppi}(2003)}]{maccarone03}
{Maccarone}, T.~J. \& {Coppi}, P.~S. 2003, \mnras, 338, 189

\bibitem[{{Markoff} {et~al.}(2005){Markoff}, {Nowak}, \& {Wilms}}]{markoff05}
{Markoff}, S., {Nowak}, M.~A., \& {Wilms}, J. 2005, \apj, 635, 1203

\bibitem[{{McClintock} \& {Remillard}(2006)}]{mclintock06}
{McClintock}, J.~E. \& {Remillard}, R.~A. 2006, {Black hole binaries}, ed.
  W.~H.~G. {Lewin} \& M.~{van der Klis}, 157--213

\bibitem[{{Migliari} \& {Fender}(2006)}]{migliari06}
{Migliari}, S. \& {Fender}, R.~P. 2006, \mnras, 366, 79

\bibitem[{{Migliari} {et~al.}(2003){Migliari}, {Fender}, {Rupen}, {Jonker},
  {Klein-Wolt}, {Hjellming}, \& {van der Klis}}]{migliari03}
{Migliari}, S., {Fender}, R.~P., {Rupen}, M., {et~al.} 2003, \mnras, 342, L67

\bibitem[{{Migliari} {et~al.}(2004){Migliari}, {Fender}, {Rupen}, {Wachter},
  {Jonker}, {Homan}, \& {van der Klis}}]{migliari04}
{Migliari}, S., {Fender}, R.~P., {Rupen}, M., {et~al.} 2004, \mnras, 351, 186

\bibitem[{{Migliari} {et~al.}(2007){Migliari}, {Miller-Jones}, {Fender},
  {Homan}, {Di Salvo}, {Rothschild}, {Rupen}, {Tomsick}, {Wijnands}, \& {van
  der Klis}}]{migliari07}
{Migliari}, S., {Miller-Jones}, J.~C.~A., {Fender}, R.~P., {et~al.} 2007, \apj,
  671, 706

\bibitem[{{Miller-Jones} {et~al.}(2010){Miller-Jones}, {Sivakoff},
  {Altamirano}, {Tudose}, {Migliari}, {Dhawan}, {Fender}, {Garrett}, {Heinz},
  {K{\"o}rding}, {Krimm}, {Linares}, {Maitra}, {Markoff}, {Paragi},
  {Remillard}, {Rupen}, {Rushton}, {Russell}, {Sarazin}, \&
  {Spencer}}]{miller-jones10}
{Miller-Jones}, J.~C.~A., {Sivakoff}, G.~R., {Altamirano}, D., {et~al.} 2010,
  \apjl, 716, L109

\bibitem[{{Mitsuda} {et~al.}(1984){Mitsuda}, {Inoue}, {Koyama}, {Makishima},
  {Matsuoka}, {Ogawara}, {Suzuki}, {Tanaka}, {Shibazaki}, \&
  {Hirano}}]{mitsuda84}
{Mitsuda}, K., {Inoue}, H., {Koyama}, K., {et~al.} 1984, \pasj, 36, 741

\bibitem[{{Mu{\~n}oz-Darias} {et~al.}(2014){Mu{\~n}oz-Darias}, {Fender},
  {Motta}, \& {Belloni}}]{munoz14}
{Mu{\~n}oz-Darias}, T., {Fender}, R.~P., {Motta}, S.~E., \& {Belloni}, T.~M.
  2014, \mnras, 443, 3270

\bibitem[{{Narayan} {et~al.}(1997){Narayan}, {Garcia}, \&
  {McClintock}}]{narayan97}
{Narayan}, R., {Garcia}, M.~R., \& {McClintock}, J.~E. 1997, \apjl, 478, L79

\bibitem[{{Narayan} \& {Yi}(1994)}]{narayan94}
{Narayan}, R. \& {Yi}, I. 1994, \apjl, 428, L13

\bibitem[{{Nowak} {et~al.}(1999){Nowak}, {Vaughan}, {Wilms}, {Dove}, \&
  {Begelman}}]{nowak99}
{Nowak}, M.~A., {Vaughan}, B.~A., {Wilms}, J., {Dove}, J.~B., \& {Begelman},
  M.~C. 1999, \apj, 510, 874

\bibitem[{{Olive} \& {Barret}(2001)}]{olive01}
{Olive}, J.-F. \& {Barret}, D. 2001, X-ray Astronomy: Stellar Endpoints, AGN,
  and the Diffuse X-ray Background, 599, 814

\bibitem[{{Oosterbroek} {et~al.}(1994){Oosterbroek}, {Lewin}, {van Paradijs},
  {van der Klis}, {Penninx}, \& {Dotani}}]{oosterbroek94}
{Oosterbroek}, T., {Lewin}, W.~H.~G., {van Paradijs}, J., {et~al.} 1994, \aap,
  281, 803

\bibitem[{{{\"O}zel} {et~al.}(2000){{\"O}zel}, {Psaltis}, \&
  {Narayan}}]{ozel00}
{{\"O}zel}, F., {Psaltis}, D., \& {Narayan}, R. 2000, \apj, 541, 234

\bibitem[{{Paizis} {et~al.}(2006){Paizis}, {Farinelli}, {Titarchuk},
  {Courvoisier}, {Bazzano}, {Beckmann}, {Frontera}, {Goldoni}, {Kuulkers},
  {Mereghetti}, {Rodriguez}, \& {Vilhu}}]{paizis06}
{Paizis}, A., {Farinelli}, R., {Titarchuk}, L., {et~al.} 2006, \aap, 459, 187

\bibitem[{{Penninx} {et~al.}(1988){Penninx}, {Lewin}, {Zijlstra}, {Mitsuda}, \&
  {van Paradijs}}]{penninx88}
{Penninx}, W., {Lewin}, W.~H.~G., {Zijlstra}, A.~A., {Mitsuda}, K., \& {van
  Paradijs}, J. 1988, \nat, 336, 146

\bibitem[{{Pottschmidt} {et~al.}(2003){Pottschmidt}, {Wilms}, {Nowak},
  {Pooley}, {Gleissner}, {Heindl}, {Smith}, {Remillard}, \&
  {Staubert}}]{pottschmidt03}
{Pottschmidt}, K., {Wilms}, J., {Nowak}, M.~A., {et~al.} 2003, \aap, 407, 1039

\bibitem[{{Poutanen} {et~al.}(1997){Poutanen}, {Krolik}, \&
  {Ryde}}]{Poutanen97}
{Poutanen}, J., {Krolik}, J.~H., \& {Ryde}, F. 1997, \mnras, 292, L21

\bibitem[{{Pozdnyakov} {et~al.}(1983){Pozdnyakov}, {Sobol}, \&
  {Syunyaev}}]{pozdnyakov83}
{Pozdnyakov}, L.~A., {Sobol}, I.~M., \& {Syunyaev}, R.~A. 1983, Astrophysics
  and Space Physics Reviews, 2, 189

\bibitem[{{Reig} \& {Kylafis}(2015)}]{reig15}
{Reig}, P. \& {Kylafis}, N. 2015, \aap, 0, in press

\bibitem[{{Reig} {et~al.}(2003){Reig}, {Kylafis}, \& {Giannios}}]{reig03}
{Reig}, P., {Kylafis}, N.~D., \& {Giannios}, D. 2003, \aap, 403, L15, "Paper I"

\bibitem[{{Revnivtsev} {et~al.}(2014){Revnivtsev}, {Tsygankov}, {Churazov}, \&
  {Krivonos}}]{revnivtsev14}
{Revnivtsev}, M.~G., {Tsygankov}, S.~S., {Churazov}, E.~M., \& {Krivonos},
  R.~A. 2014, \mnras, 445, 1205

\bibitem[{{Shakura} \& {Sunyaev}(1973)}]{shakura73}
{Shakura}, N.~I. \& {Sunyaev}, R.~A. 1973, \aap, 24, 337

\bibitem[{{van der Klis}(1994)}]{vanderklis94}
{van der Klis}, M. 1994, \apjs, 92, 511

\bibitem[{{van der Klis}(2005)}]{vanderklis05}
{van der Klis}, M. 2005, \apss, 300, 149

\bibitem[{{van der Klis}(2006)}]{vanderklis06}
{van der Klis}, M. 2006, {Rapid X-ray Variability}, ed. W.~H.~G. {Lewin} \&
  M.~{van der Klis}, 39--112

\bibitem[{{Zdziarski} {et~al.}(2001){Zdziarski}, {Grove}, {Poutanen}, {Rao}, \&
  {Vadawale}}]{zdziarski01}
{Zdziarski}, A.~A., {Grove}, J.~E., {Poutanen}, J., {Rao}, A.~R., \&
  {Vadawale}, S.~V. 2001, \apjl, 554, L45

\end{thebibliography}

\end{document}